\documentclass[aps,showpacs,amsmath,amssymb,nofootinbib,preprintnumbers]{revtex4}

\usepackage{graphicx}
\usepackage{color}

\def \be  {\begin{equation}}
\def \ee  {\end{equation}}
\def \bea {\begin{eqnarray}}
\def \eea {\end{eqnarray}}

\def \Tr  {\bf{Tr}}

\begin{document}

\preprint{ECTP-2010-06}

\title{Dynamical Fluctuations in Baryon--Meson Ratios}

\author{A.~Tawfik}
\email{drtawfik@mti.edu.eg}
\affiliation{Egyptian Center for Theoretical Physics (ECTP), MTI University,
 Cairo-Egypt}

\date{\today}

\begin{abstract}
The event--by--event dynamical fluctuations in kaon--to--proton and proton--to--pion ratios have been studied in dependence on center--of--mass energies of nucleon--nucleon collisions $\sqrt{s}$. Based on changing phase space volume which apparently is the consequence of phase transition from hadrons to quark--gluon plasma at large $\sqrt{s}$, the single--particle distribution function $f$ is assumed to be rather modified. 
Varying $f$ and phase space volume are implemented in the grand--canonical partition function, especially at $\sqrt{s}>17$ GeV, so that hadron resonance gas model, when taking into account the experimental acceptance ${\cal A}$ and quark phase space occupation factor $\gamma$, turns to be able to reproduce the dynamical fluctuations in $(K^++K^-)/(p+\overline{p})$ and $(p+\overline{p})/(\pi^++\pi^-)$  ratios over the entire range of $\sqrt{s}$.  
\end{abstract}


\maketitle

\section{Introduction}

In order to characterize the fluctuations and collective properties in heavy--ion collisions, phase structure and event--by--event fluctuations have been suggested ~\cite{tawPS1,shury,rajag,koch}. 
From {\it theoretical} point--of--view, the dynamical fluctuations are particularly essential regarding to examining of existing statistical models \cite{giorg,Karsch:2003vd}, characterizing of particle equilibration \cite{pequil} and the search for unambiguous signals of {\it new state of matter} \cite{qgpA,qgpB}. From the {\it phenomenological} point--of--view, they have been measured at SPS and RHIC energies~\cite{Roland,NA491,STAR}, i.e, $6<\sqrt{s}<200$ GeV. 

The success of hadron resonance gas model (HRG) in describing the dynamical fluctuations in kaon--to--pion ratios \cite{tawPS1} and the ingredient to have other implications are the motivations of this work, where the fluctuations in the remaining particle ratios, that have been measured so far, i.e, kaon--to--proton and proton--to--pion, shall be analyzed. 
To bring theoretical and experimental data very close to each others, the experimental acceptance ${\cal A}$, grand--canonical statistics and quark phase space occupation factor, $\gamma$, have been utilized. It has been noticed that the fluctuations over the whole range of center--of--mass energies $\sqrt{s}$ exhibit a non-monotonic behavior and a certain configuration of $\gamma$, ${\cal A}$ and grand--canonical statistics, are not able to reproduce the whole experimental data.     

The concept of phase space dominance suggested by Fermi \cite{fermii} six decades ago is applied in this model. The assumption of {\it equilibrium} single--particle distribution function $f$ and {extensive} thermodynamics that have been shown to perfectly reproduce all thermodynamic quantities, including multiplicities and fluctuations, as long as the energy density is not high enough to derive the hadronic system into the {\it new state of matter}, quark--gluon plasma (QGP) should be no longer valid, particularly, when the energy density exceeds the critical value, as the case at RHIC and LHC energies. Across the phase transition, the phase space volume, symmetries and accordingly the effective degrees of freedom are likely subject of a prompt change. Therefore, the phase space volume, in which the microscopic states are distributed according to $f$ function is conjectured to be a subject of modification, as well. In present work, changing phase space volume and $f$ are implemented in the grand--canonical partition function, explicitly at $\sqrt{s}>17$ GeV, from which the particle number and fluctuation have been calculated.

There are many reasons to apply HRG model. This model provides a good description for the thermodynamical evolution of the hadronic system below the critical temperature~\cite{Karsch:2003vd,Karsch:2003zq,Redlich:2004gp} and has been successfully used to characterize the conditions deriving the chemical freeze-out~\cite{Taw3b,Taw3c}. 
Apparently, there are many reasons speak for studying of dynamical fluctuations in the baryon--meson yield ratios. {\it Statistically}, the interplay between fermions and bosons is crucial. The energy threshold required to create baryons and mesons is of a great interest. Size of the system is another ingredient of this study. {\it Phenomenologically}, the strange fluctuations, especially, when passing through {\it deconfinement} phase transition, are conjectured to affect the dynamical fluctuations in the hadronic final state. The strange dynamical fluctuations are expected to survive through the mixed phase. Therefore, these particle ratios are sensitive to the symmetry change and the dynamics of deconfinement and chiral phase transitions, respectively.

The paper is organized as follows. The model is given in section \ref{sec:model}, where non--extensive single--particle equilibrium distributions are introduced. Section \ref{sec:dynmFlct} is devoted to the dynamical fluctuations in kaon--to--proton and proton--to--pion ratios. Discussion and final conclusions are elaborated in section \ref{sec:concls}.

\section{Model}
\label{sec:model}

\subsection{Single--Particle Equilibrium Distribution of Hadrons}

The grand-canonical partition function is given by Hamiltonian and baryon number operators, $\hat{H}$ and $\hat{b}$, respectively, and depends on temperature $T=1/\beta$ and chemical potential $\mu$, 
\bea \label{eq:zTr}
Z(\beta,V,\mu) &=& \Tr \, \left[\gamma\, \exp^{\beta (\mu \hat{b}-\hat{H})}\right].
\eea
It can be characterized by various but a complete set of microscopic
states and therefore the physical properties of the quantum systems turn to be accessible in approximation of non-correlated {\it free} hadron resonances. Therefore, the resonances are treated as a free gas~\cite{Karsch:2003vd,Karsch:2003zq,Redlich:2004gp,Tawfik:2004sw,Taw3}. Each of them is  conjectured to add to the overall thermodynamic pressure of the {\it strongly} interacting hadronic matter. This statement is valid for free, as well as, strong interactions. It has been shown that the thermodynamics of strongly interacting system can be approximated to an ideal gas composed of hadron resonances with masses $\le 2~$GeV~\cite{Tawfik:2004sw,Vunog}. The heavier constituents, the smaller thermodynamical quantities. The main motivation of using the Hamiltonian is that it contains all relevant degrees of freedom of confined and strongly interacting matter. It implicitly includes the interactions that result in the formation of new resonances. In addition, this model has been shown to provide a quite satisfactory description of the particle production and collective properties in heavy--ion collisions. 

The conservation of baryon number $n_b$ represents an additional constrain on grand--canonical partition function. Therefore, Eq. (\ref{eq:zTr}), can be re--written as  
\bea \label{eq:zTr-Lagrange}
Z_{gc}(T,V,\mu) &=& \Tr \, \left[\exp^{\frac{\mu \hat{b}-\hat{H}}{T}-\alpha}\right],\\
f_{gc}(T,V,\mu) &=& \frac{\exp\left(\frac{-\hat{H}}{T}-\alpha\right)}{Z_{gc}(T,V,\mu)},
\eea
where $\alpha$, besides $\beta$, are Lagrange multipliers in the entropy maximization. The physical meaning of $\alpha$ is controller over number of particles in the phase space, i.e, acting as chemical potential. 
\bea
\alpha &=&  \ln {\cal E} - \ln T - \ln N.
\eea
It combines intensive variables, $T$ and $N$ with an extensive one ${\cal E}=\sum_i^n g_i\epsilon_i$, where $g_i$ and $\epsilon_i$ are degeneracy factor and energy of $i$--th cell in the phase space, respectively. 
The most probable state density is to be found by Lagrange multipliers, where one of them, $\alpha$, has been expressed in term of the second one, $\beta$, and the occupation numbers of the system. Apparently, $\alpha$ gives how the energy ${\cal E}$ is distributed in the microstates of the equilibrium system and therefore, can be understood as another factor controlling the number of occupied states, at the microcanonical level.

Under these assumptions, the dynamics of partition function can be calculated as a summation over {\it single--particle} partition functions $Z_{gc}^i$ of all hadrons resonances.
\bea
\ln Z_{gc}(T,V,\mu)&=&\sum_i \ln Z_{gc}^i(T,V,\mu)=\sum_i\pm \frac{g_i}{2\pi^2}\,V\int_0^{\infty} k^2 dk \ln\left(1\pm \gamma\,\lambda_i\, e^{-\epsilon_i(k)/T}\right)
\eea
where $\pm$ stands for bosons and fermions, respectively.  $\lambda_i=\exp(\mu_i/T)$ is the $i$-th particle fugacity. As given above, $\gamma=\exp(-\alpha)$ is the quark phase space occupation factor.

As obtained in Ref. \cite{tawPS}, the {\it equilibrium} distribution function $f$ is conjectured to be no longer valid, especially, when the energy density is high enough to derive {\it confined} hadronic into {\it deconfined} partonic matter. The equilibrium is settled, when $\partial_{;\,t}\,S=0$ and the probability current entirely vanishes, as well \cite{3ppBb,crrent}.    
\bea
f(\vec{x},\vec{k},t) &\simeq& f_{eq}(\vec{x},\vec{k})\; {\cal Q}(\vec{x},\vec{k}), \label{eq:nf}
\eea
where ${\cal Q}(\vec{x},\vec{k})$ is conjectured to reflect the change in phase--space when the hadronic degrees of freedom are replaced by partonic ones. It can be interpreted as a measure for the non--extensivity. 
Therefore, the partition functions reads,
\bea
\ln Z_{gc}(T,{\cal V},\mu)&=& \sum_i\pm \frac{g_i}{2\pi^2}\,{\cal V}\int_0^{\infty} k^2 dk\; \ln\,\left(1\pm \gamma\,{\cal Q}\, \exp\left[\frac{\mu_i-\epsilon_i(k)}{T}\right]\right),
\eea
where single--particle distribution function and the particle number can be written as 
\bea \label{eq:nfinal2}
f_{gc}(\vec{k}, \mu) 
   &=& \left\{ \gamma^{-1}{\cal Q}^{-1}(\vec{k})\exp\left[\frac{\epsilon(\vec{k})-\mu}{T}\right] \pm 1\right\}^{-1},\\
\label{eq:nNonEqu}
n(T,{\cal V},\mu) &=& \frac{\partial}{\partial \mu}\, \lim_{{\cal V}\rightarrow \infty} \frac{T}{{\cal V}} \ln Z_{gc}(T,{\cal V},\mu).
\eea

\section{Dynamical Fluctuations in Baryon--Meson Ratios}
\label{sec:dynmFlct}

The fluctuations in particle number, Eq. (\ref{eq:nNonEqu}), are mainly given by the susceptibility, which is the derivative of particle number $\langle n\rangle$ wrt chemical potential $\mu$.  
\bea 
\label{eq:n1} 
\langle n\rangle &=& \sum_i \frac{g_i}{2\pi^2} \int_0^{\infty} k^2 dk 
\frac{\gamma\,{\cal Q}}{\exp\left[\frac{\epsilon_i(k)-\mu_i}{T}\right] \pm \gamma\,{\cal Q}}, \\
\label{eq:dn1} 
\langle (\Delta n)^2\rangle &=& \sum_i \frac{g_i}{2\pi^2} \frac{1}{T} \int_0^{\infty} k^2 dk 
           \frac{\gamma\,{\cal Q}\,\exp\left[\frac{\epsilon_i(k)-\mu_i}{T}\right]}
	   {\left(\exp\left[\frac{\epsilon_i(k)-\mu_i}{T}\right] \pm \gamma\,{\cal Q}\right)^2},
\eea
where $\langle\cdots\rangle$ stands for statistical average.

At final state, i.e, when the {\it chemically relaxing} system absolves the chemical freeze-out process, the resonances are conjectured to decay either to stable particles or
to other resonances. This chemical process has to be take into account in the particle numbers and fluctuations given above as follows.
\bea \label{eq:n2}
\langle n_i^{final}\rangle &=& \langle n_i^{direct}\rangle + \sum_{j\neq
i} b_{j\rightarrow i} \langle n_j\rangle,\\ \label{eq:dn2} 
\langle (\Delta n_{j\rightarrow i})^2\rangle &=& b_{j\rightarrow i} (1-b_{j\rightarrow
i}) \langle n_j\rangle + b_{j\rightarrow i}^2 \langle (\Delta
n_{j})^2\rangle, 
\eea
where $b_{j\rightarrow i}$ is the  branching ratio for the decay of $j$-th resonance
to $i$-th particle. In order to characterize the stage at which the chemical freeze--out takes place, we use the model introduced in Ref. \cite{Taw3b,Taw3c,Taw3}. The ratio $s/T^3$, where $s=S/V$ is the entropy density, is assigned to a constant value. This thermodynamical condition perfectly reproduces the freeze--out line over a wide range of chemical potential, $\mu$, which in turn can be {\it empirically} expressed in center--of-mass energy, $\sqrt{s}$ in GeV, \cite{sqrts_mu}
\bea
\sqrt{s} &=& \frac{1}{0.614} \left(\frac{3.286}{\mu}-2.53\right)
\eea  

The event--by--event fluctuations in a ratio of two particles $n_1/n_2$ are~\cite{koch}
\bea  \label{eq:sigma}
\sigma^2_{n_1/n_2} &=& \frac{\langle (\Delta n_1)^2\rangle}{\langle n_1\rangle^2} + 
                       \frac{\langle (\Delta n_2)^2\rangle}{\langle n_2\rangle^2} - 
                     2 \frac{\langle \Delta n_1 \; \Delta
		     n_2\rangle}{\langle n_1\rangle \; \langle
		     n_2\rangle},
\eea
which includes both dynamical and statistical fluctuations. The third term of
Eq.~\ref{eq:sigma} counts for fluctuations from the hadron resonances which 
decay into particle $1$ and particle $2$, simultaneously. In such a
mixing channel, all correlations including quantum statistics ones are
taken into account. Obviously, 
this decay channel results in strong correlated particles. To extract
statistical fluctuation, we apply Poisson scaling in mixed decay
channels. Experimentally, there are various methods to
construct statistical fluctuations~\cite{STAR}. Frequently used
method is the one that measures particle ratios from mixing events.
\bea \label{eq:sigmaStat}
(\sigma^2_{n_1/n_2})_{stat} &=& \frac{1}{\langle n_1\rangle} +
\frac{1}{\langle n_2\rangle}. 
\eea
Subtracting Eq.~\ref{eq:sigmaStat} from Eq.~\ref{eq:sigma}, then 
the dynamical fluctuations in $n_1/n_2$ ratio read
\bea
 \label{eq:sigma2}
(\sigma^2_{n_1/n_2})_{dyn} &=& 
          \frac{\langle n_1^2\rangle}{\langle n_1\rangle^2} +
          \frac{\langle n_2^2\rangle}{\langle n_2\rangle^2} -
         \frac{\langle n_1\rangle+\langle n_2\rangle +
	 2\langle n_1 n_2\rangle}{\langle n_1\rangle\langle n_2\rangle}
\eea

\section{Discussion and Conclusions}
\label{sec:concls}

\begin{figure}[htb]
\includegraphics[angle=-90,width=8.5cm]{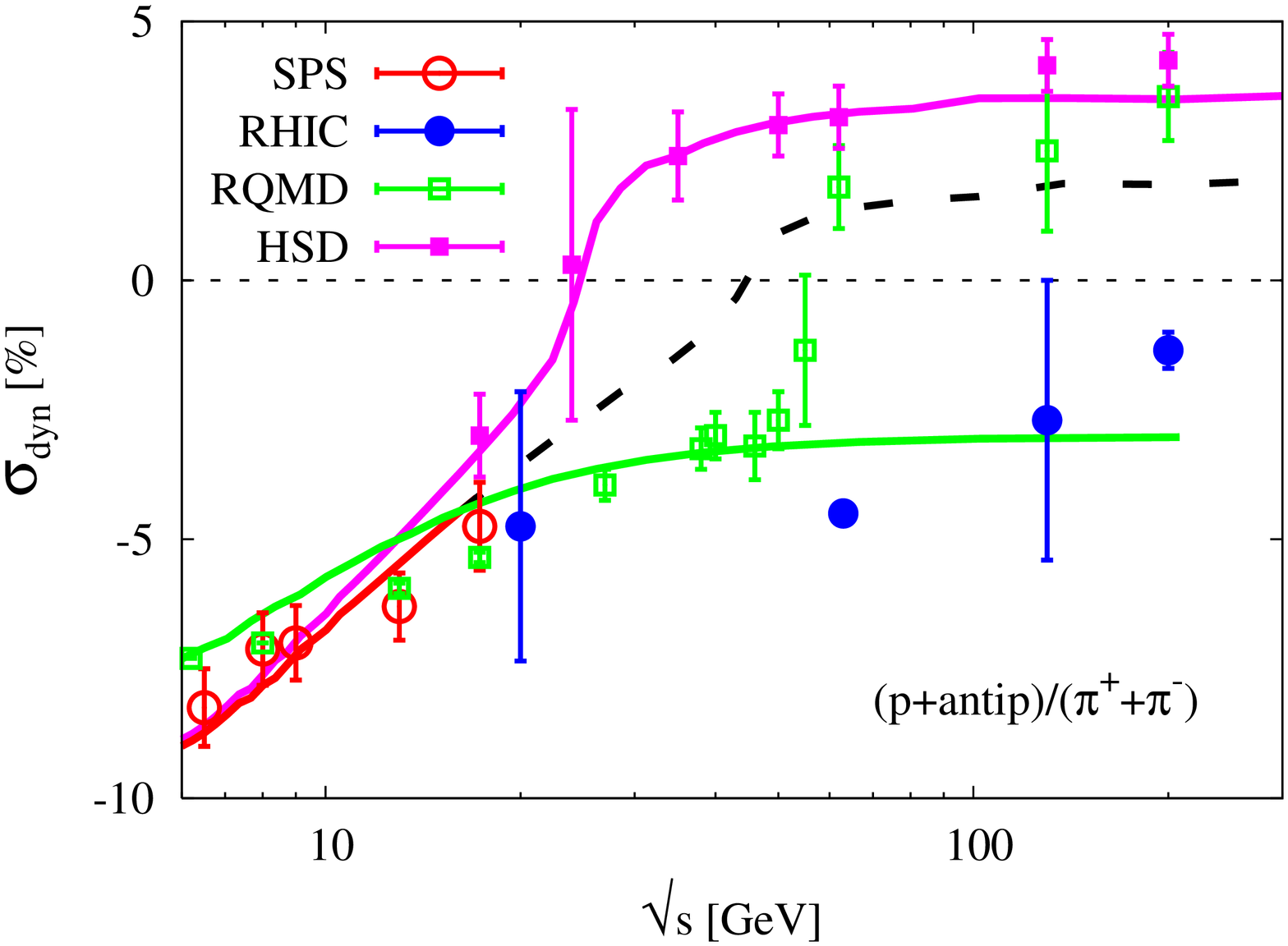}
\includegraphics[angle=-90,width=8.5cm]{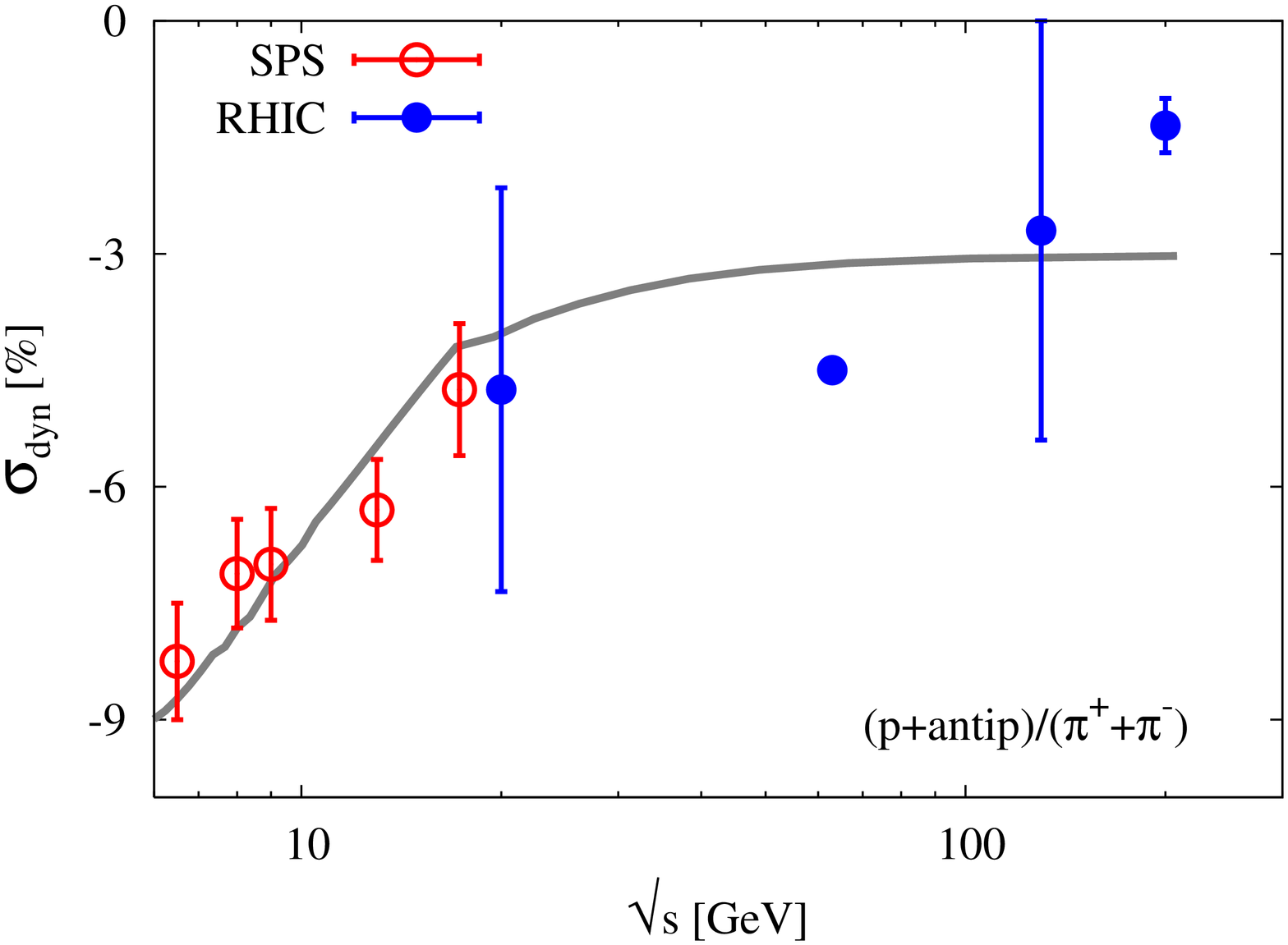}
\caption{\normalsize Dynamical fluctuations in $(p+\overline{p})/(\pi^++\pi^-)$ ratio as function of center--of--mass energy $\sqrt{s}$. Taking into account the experimental acceptance ${\cal A}$, the lower curve represents HRG results at finite value for $\gamma$. It reproduces perfectly the SPS data. The dashed region shows that RHIC data are largely underestimated. The upper curve represents HRG that reproduces HSD simulations. RQMD and HSD have been proceeded using RHIC configurations. When phase space factor ${\cal Q}$ is switched on, HRG matches with RHIC data. The overall curve is given in the right panel.} 
\label{fig:1a1}
\end{figure}

\begin{figure}[htb]
\includegraphics[angle=-90,width=8.5cm]{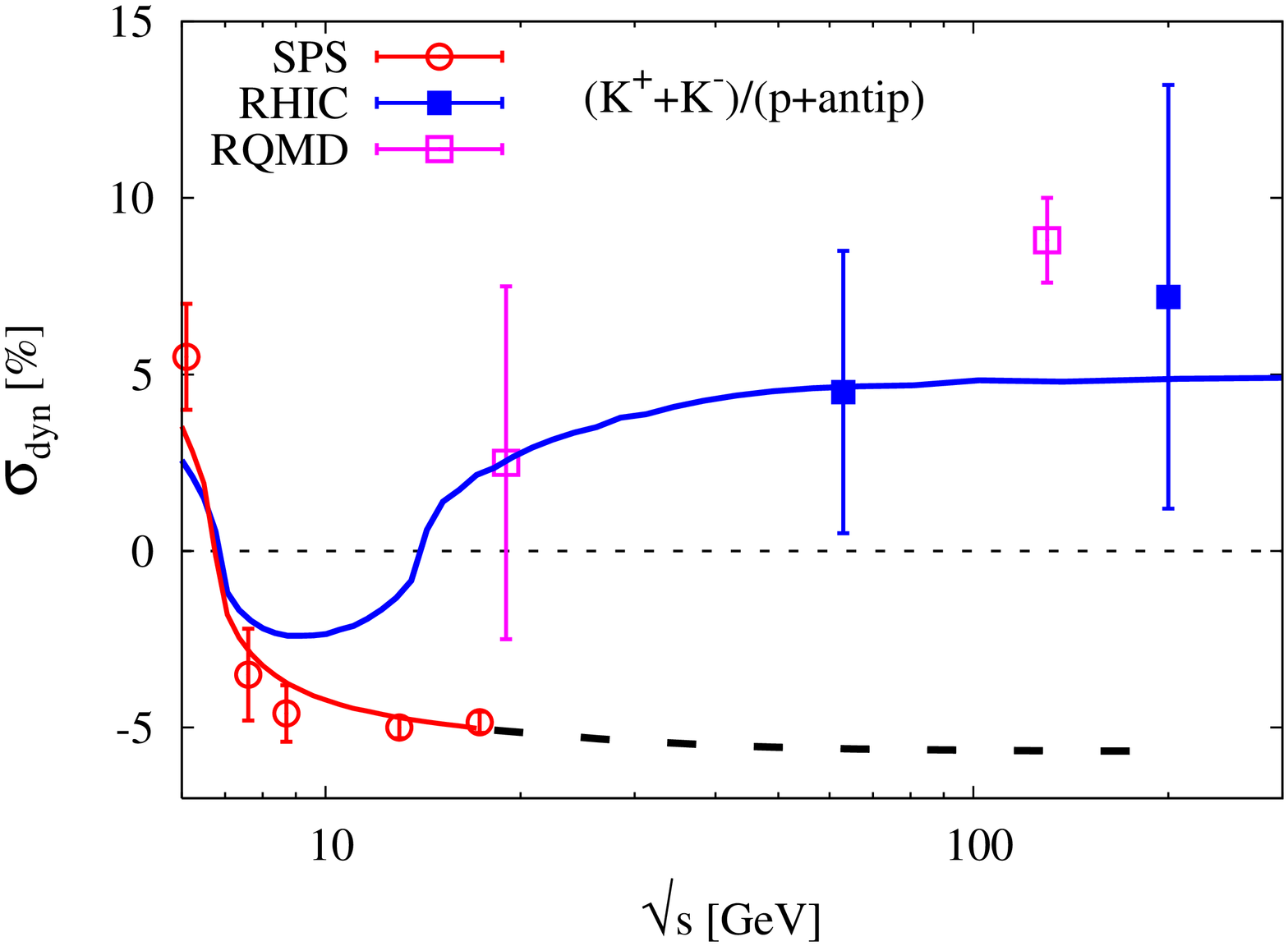}
\includegraphics[angle=-90,width=8.5cm]{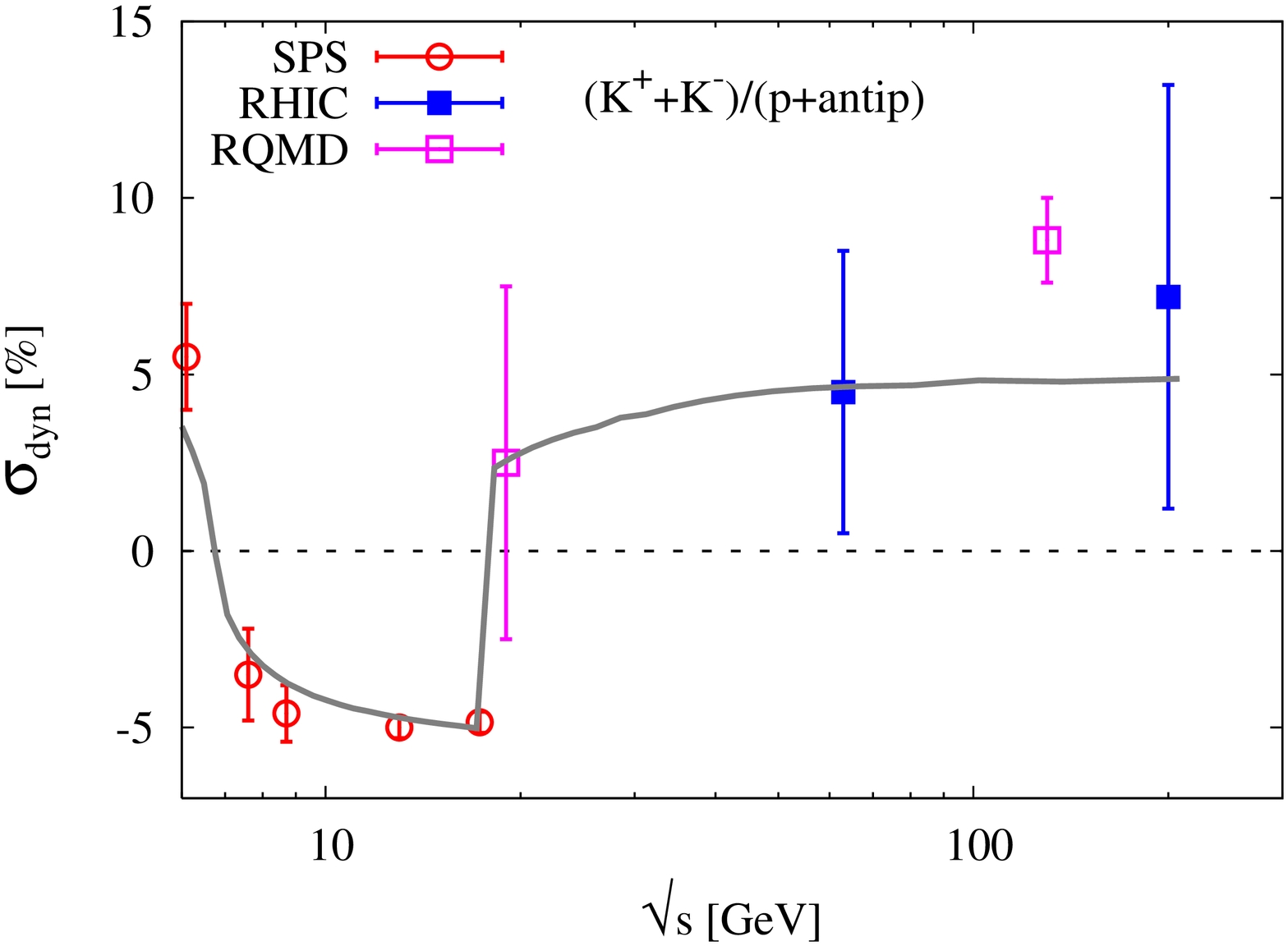}
\caption{\normalsize Dynamical fluctuations in $(K^++K^-)/(p+\overline{p})$ ratio as function of center--of--mass energy $\sqrt{s}$. SPS fluctuations rapidly decrease to negative values, while RHIC fluctuations are positive and slowly increase with increasing $\sqrt{s}$. Taking into account the experimental acceptance ${\cal A}$, the lower curve represents HRG results at finite value for $\gamma$, that satisfactorily match with SPS data. The dashed region shows that RHIC data are largely underestimated. The upper curve represents HRG results, when phase space factor ${\cal Q}$ is switched on. RHIC data perfectly matches with this curve. The curves are not stemming from statistical fitting process.} 
\label{fig:2a1}
\end{figure}

As introduced previously, HRG assumes that the hadron resonances are point-like and non--correlated free gas. Therefore, it is conjectured that the average multiplicity $\langle n\rangle$, Eq.~(\ref{eq:n2}), and the dynamical fluctuations $\sigma^2_{N_1/N_2}$, Eq.~(\ref{eq:sigma}), are not strongly dependent on the volume fluctuations. Therefore, we assume that volume fluctuations over the entire range of $\sqrt{s}$ are minimum and thus neglected. The experimental acceptances of the different detectors ${\cal A}$ have been taken into account. Also, the quark phase space occupation factor, $\gamma$, has been estimated.  The dynamical fluctuations in particle yield ratios with and without ${\cal Q}$ are plotted in Figs. \ref{fig:1a1} \& \ref{fig:2a1}.

Left panels in both figures show a full comparison between different data using  different parameters. Right panels give the final results that satisfactorily describe the non--monotonic behavior. The quark phase space occupation factor $\gamma$ {\it alone} is apparently not able to reproduce the whole data set, although the experimental acceptances of the different detectors have been taken into account. The largest disagreement occurs at $\sqrt{s}>17\,$GeV, is given by bottom curves, especially the dashed parts. RQMD and HSD are used to generate dynamical fluctuations in corresponding particle yield ratios using RHIC configurations. Therefore, their agreement with RHIC data is satisfactory. The dynamical fluctuations in same particle yield ratios and strangeness production in HSD transport approach from SIS and SPS energies has been given in Ref. \cite{hsd1}. Although, both RQMD and HSD reflect a global behavior that the fluctuations alternate between negative and positive values, RQMD seems to work nicely at very high energies while HSD obviously covers a much wide range of energy. 

Some details about RQMD and HSD are now in order. The relativistic quantum molecular dynamics (RQMD) is based on color rope mechanism and excitation and/or fragmentation of color strings followed by hadronic re--interactions. 
The color rope mechanism represents an important source for the production of  strange hadrons. The comparison with experimental data shows that RQMD reproduces very well the main properties in hadron production and collective properties. Therefore, it is believed to be a reliable model for generating the full momentum space of hadron resonances. The particle yields from RQMD would be accompanied with statistical uncertainties that basically can be kept arbitrarily small, i.e, generation of large ensemble. That the microscopic state of RQMD at chemical freeze-out is in good agreement with chemical and kinetic
equilibrium leads to believe in RQMD chemical freeze-out parameters \cite{rqmd}.

The HSD transport approach is based on various degrees of freedom, including quark, diquark, string and hadron. It describes the creation of dense/hot hadronic matter, the in--medium modification of hadron properties and the overall dynamics.
HSD gives {\it numerical} solution of a coupled set of relativistic transport equations for particles with in--medium self--energies of test--particle. It applies two approaches. One at high energy, where the inelastic interactions are described by MC--techniques, FRITIOF string model. At low energy, modeling of hadron collisions is based on experimental inputs. The transport approach implemented in it takes into account formation and multiple re--scattering of leading pre-hadrons and hadrons and has been designed to reproduce nucleon-nucleon, meson-nucleon and meson-meson cross section data in a wide kinematic range \cite{hsd1}.

In addition to the assumptions introduced in \cite{kpi_horn,kochSchuster}, we give here a novel one. We assume that the prompt raise at $\sqrt{s}\sim17$ GeV is to be understood according to a modification in the phase space volume. To this end, it has been concluded in Ref. \cite{tawPS1} that $f$ is a subject of modification, especially at large $\sqrt{s}$. In this limit, the energy density available to the system turns to be high enough to cause the hadronic matter, where equilibrium $f$ is perfectly able to reproduce almost all essential transport properties and thermodynamic quantities, including the dynamical fluctuations, to go through a phase transition into QGP. Such a phase transition apparently results in various types of modifications, such as symmetries and degrees of freedom. Also the configurations of microstates in phase space volume $d^3\vec{x}d^3\vec{k}$ and the single--particle distribution function $f$ are not exceptions.   
 
Here, we apply this model to the dynamical fluctuations in $(K^++K^-)/(p+\overline{p})$ and $(p+\overline{p})/(\pi^++\pi^-)$ ratios. These particle yield ratios combine both strangeness and light boson fluctuations. Additionally, they are very sensitive to the deconfinement and chiral phase transitions, respectively. RQMD and HSD simulations result is positive fluctuations at high energy. Implementing $\gamma$ and ${\cal Q}$ in the grand canonical partition function of HRG results in a very well description of the experimentally measured fluctuations in these particle yields over the entire range of $\sqrt{s}$. Although the experimental fluctuations are negative overall, the non-monotonic behavior is very well reproduced over a wide range of $\sqrt{s}$. It is obvious, that these dynamical fluctuations non--avoidablely refer to non-extensive and non--equilibrium state of matter, that basically differs from the one at SPS energies. The modification of the state of matter has been combined in the factor ${\cal Q}(\vec{x},\vec{k})\simeq\exp(2\omega\,\vec{x}\,\vec{k})$ whose numerical value is given in Tab. \ref{tab:1a1} and kept unchanged with changing $\sqrt{s}$, right panel of Figs. \ref{fig:1a1} and \ref{fig:2a1}.  

Left panel of Fig. \ref{fig:1a1} includes SPS and RHIC experimental data (circles) with RQMD and HSD simulations (rectangles) for the dynamical fluctuations in $(K^++K^-)/(p+\overline{p})$ ratio. The results from HRG model are given by the curves. It is clear that SPS data can be reproduced by HRG using a certain set of $\gamma$ and ${\cal A}$ parameters. The whole set of parameters is given in Tab. \ref{tab:1a1}. The parameters suitable for SPS are no longer able to reproduce RHIC data. To this end, ${\cal Q}$ has to be switched on. It is clear that HSD simulations can be reproduced in HRG, while RQMD not overall, especially in middle region. Almost the same behavior is present in the left panel of Fig. \ref{fig:2a1} for $(p+\overline{p})/(\pi^++\pi^-)$ ratio. Here, the fluctuations are alternating between positive and negative values. Even at low energy, $\sigma_{dyn}$ decreases and flips its positive sign, with increasing $\sqrt{s}$. At $\sqrt{s}>17$ GeV, $\sigma_{dyn}$ suddenly jumps into the positive region. This prompt sign--exchange can be taken as an order parameter. When the energy available to the system is high enough to create protons, the dynamical fluctuations get positive. The parameters $\gamma$ and ${\cal A}$ that have been used to reproduce SPS data turn to underestimate the RHIC results, the dashed curve. Implementing ${\cal Q}$ produces the top curve, which matches satisfactorily with RHIC results but not with SPS, although the interesting behavior, at low energy.

We conclude that HRG model reproduces the experimentally measured fluctuations in $(K^++K^-)/(p+\overline{p})$ and $(p+\overline{p})/(\pi^++\pi^-)$  ratios over the entire range of $\sqrt{s}$. In generating this excellent agreement no statistical fitting has been performed. Two essential parameters have been merely adjusted with the experimental acceptance in order to reproduce the experimental data. 

In light of the best reproduction of dynamical fluctuations that have been measured in various particle ratios over a wide range of $\sqrt{s}$, this model, which has been introduced in Ref. \cite{tawPS1}, would be a suitable platform to describe other collective properties, such as fluctuations in nett charge, and energy, etc. Other implications are also possible. Based on the pioneering work about bulk and shear viscosity in hadronic matter \cite{tawBS1}, the transport properties turn to be accessible by means of of this model. Thus, the comparison with the transport models, like RQMD, UrQMD and HSD, etc. shall get more and more creditability.       

\begin{table*}
	\begin{center}
		\begin{tabular}{|| c | c || c | c | c ||}\hline\hline
      Particle Ratio   & Data Set    & Occupation $\gamma$ & Acceptance ${\cal A}$ & Phase space ${\cal Q}$ \\ \hline\hline
         & SPS & $0.35$ & $0.1$ & $-$   \\ 
 $K^++K^-)/(p+\overline{p})$   & RHIC&  $0.35$ & $0.1$  & $2.285$ \\ 
         & RQMD& $0.8$ & $0.1$  & $-$  \\ \hline\hline
         & SPS & $0.55$ & $0.55$ & $-$   \\ 
 $(p+\overline{p})/(\pi^++\pi^-)$   & RHIC&  $0.55$ & $0.55$  & $0.636$ \\ 
         & HSD & $0.6$ & $0.55$  & $-$  \\ \hline\hline
\end{tabular}
	\end{center}
	\caption{Parameters used in HRG model in order to reproduce the experimental and simulation data}
	\label{tab:1a1}
\end{table*}
   


\end{document}